\documentclass[letterpaper, 10 pt, conference]{ieeeconf}  

\IEEEoverridecommandlockouts                              

\usepackage{graphicx} 
\usepackage[english]{babel}
\usepackage[utf8x]{inputenc}
\usepackage{graphicx}
\usepackage{amsfonts}

\usepackage{amsmath}
\usepackage{amssymb}
\usepackage{array}
\usepackage{xcolor}
\usepackage[ruled,vlined]{algorithm2e}
\usepackage{subcaption}
\usepackage{multicol}
\usepackage{hyperref}
\usepackage{url}
\usepackage{footmisc}
\usepackage{siunitx}

\newcommand{\R}{\mathbb{R}}

\newcommand{\E}{\mathbb{E}}

\newcommand{\bx}{\mathbf{x}}
\newcommand{\bv}{\mathbf{v}}
\newcommand{\bt}{\mathbf{t}}

\newcommand{\td}{\dot{\theta}}

\newcommand*\diff{\mathop{}\!\mathrm{d}}

\newcommand{\Var}{\mathrm{Var}}

\renewcommand{\epsilon}{\varepsilon}

\usepackage[]{algorithm2e}
\usepackage{ragged2e}
\usepackage{etoolbox}
\usepackage{empheq}

\title{\LARGE \bf
Risk-Sensitive Rendezvous Algorithm for Heterogeneous Agents in Urban Environments*
}

\author{Gabriel Barsi Haberfeld$^{1}$, Aditya Gahlawat$^{2}$, and Naira Hovakimyan$^{3}$
\thanks{*This work was supported by NSF NRI award 1830639.}
\thanks{$^{1}$Gabriel Barsi Haberfeld is a Ph.D. Student with the Department of Mechanical Science and Engineering, University of Illinois at Urbana-Champaign
{\tt\small gbh2@illinois.edu}.}%
\thanks{$^{2}$Aditya Gahlawat is a Postdoctoral Researcher with the Department of Mechanical Science and Engineering, University of Illinois at Urbana-Champaign
{\tt\small gahlawat@illinois.edu}.}%
\thanks{$^{3}$Naira Hovakimyan is with Faculty at the Department of Mechanical Science and Engineering, University of Illinois at Urbana-Champaign
{\tt\small nhovakim@illinois.edu}.}%
}

\begin{document}

\maketitle
\thispagestyle{empty}
\pagestyle{empty}

\begin{abstract}

Demand for fast and inexpensive parcel deliveries in urban environments has risen considerably in recent years. A framework is envisioned to enforce efficient last-mile delivery in urban environments by leveraging a network of ride-sharing vehicles, where Unmanned Aerial Systems (UASs) drop packages on said vehicles, which then cover the majority of the distance before final aerial delivery. This approach presents many engineering challenges, including the safe rendezvous of both agents: the UAS and the human-operated ground vehicle. This paper introduces a framework to minimize the risk of failure while allowing for the controlled agent's optimal usage. We discuss the downfalls of traditional approaches and formulate a fast, compact planner to drive a UAS to a passive ground vehicle with inexact behavior. To account for uncertainty, we learn driver behavior while leveraging historical data, and a Model Predictive Controller minimizes a risk-enabled cost function. The resulting algorithm is shown to be fast and implementable in real-time in qualitative scenarios.

\end{abstract}

\section{INTRODUCTION}

\begin{figure}
        \centering
        \begin{subfigure}[b]{0.26325\textwidth}
            \centering
            \includegraphics[width=\textwidth]{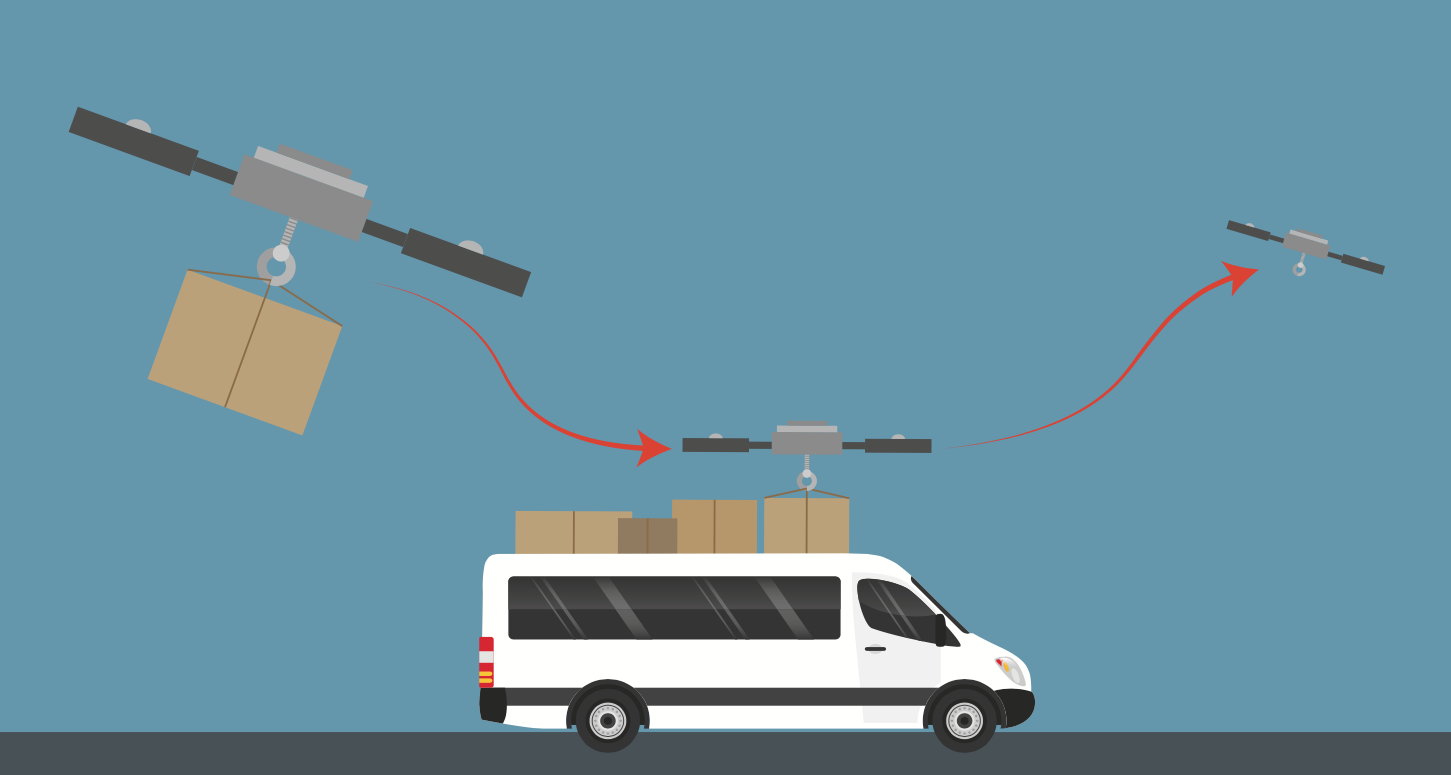}
            \caption[]%
            {{}}    
            \label{fig:proc}
        \end{subfigure}
        \begin{subfigure}[b]{0.16875\textwidth}  
            \centering 
            \includegraphics[width=\textwidth]{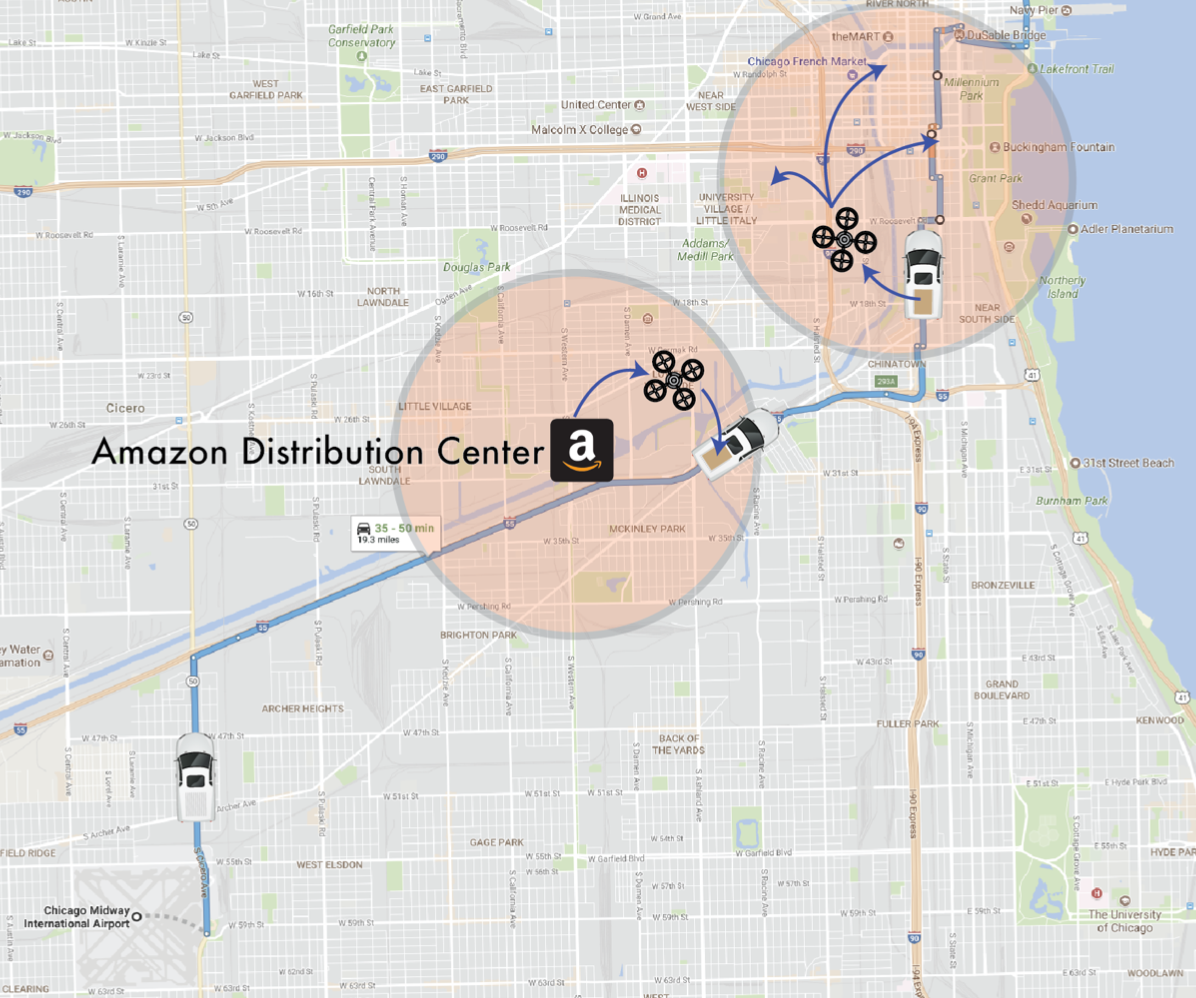}
            \caption[]%
            {{}}    
            \label{fig:longh}
        \end{subfigure}
        \caption[]
        {\small Air-ground rendezvous procedure. Left: the UAS needs to meet an uncontrollable ground vehicle with uncertain trajectory. Right: UASs intercept the vehicle at various points to complete the delivery. In this example an Amazon package is carried by a ride-sharing vehicle departing from Chicago Midway Airport bound to Downtown Chiacgo.} 
        \label{fig:introfig}
\end{figure}

Modern shipping solutions can accumulate more than half of the total shipping cost on the transportation portion between the final distribution center and the customer \cite{mckinsey}. This is known as the {\em last-mile problem}. Our proposed framework consists of using the existing large networks of ride-sharing services (Uber, Lyft) to cover most of the distance from the final distribution center to the customer. This process uses knowledge of these vehicles' destination to plan deliveries, where a UAS carries the parcel from the distribution center and places it on a moving vehicle, or picks up a package from a moving vehicle and delivers it to a final location. An example scenario is illustrated in Fig.~\ref{fig:introfig}. The critical concern is the  driver behavior. An erratic driver adds an undesirable risk to the two stages of the mission: (1) landing safely on the moving vehicle to drop the parcel and (2) flying back to the distribution center. Environmental factors such as wind, package mass, sloshing of package contents, battery age, and others contribute to these safety concerns. However, because of the long planning horizons associated with these missions, the primary source of risk and uncertainty arises from the inexact driver behavior, where a driver might be slower, faster, or erratic.

Model Predictive Control (MPC) is a  popular method for solving local Optimal Control Problems (OCP) in real-time \cite{mesbah}, where the OCP is solved at each control loop step. Although versatile, traditional MPC is not equipped to deal with large uncertainties over long planning horizons due to exponentially increasing uncertainty propagation in the planning stage. To address these issues two common solutions are (a) stochastic MPC (SMPC) \cite{stochasticmpc,primbs} and (b) Robust MPC \cite{robustmpc,kothare,yang,park}. Stochastic MPC is often referred to as \emph{risk-neutral}, as it aims to solely minimize expectations, while Robust MPC accounts for worst case scenarios. In some cases, an absolute approach is desirable, but often the problem requires a trade-off between high risk and robustness, as not to diverge too far away from optimality. In most cases, optimizing over risk measures turns the OCP intractable for real-time implementation due to the added complexity \cite{Schildbach_2014}. We solve this issue by formulating the mission so that we only need to check the risk measure constraint once.

We focus on the high-level problem of trajectory planning of a UAV to reach a neighborhood of the ground vehicle and flying to the desired landing location, where we assume that successful take-off, rendezvous, and landing are always achievable by a local, low-level controller. We address tractability by formulating the problem with a concise risk measure directly related to mission success checked only once during the mission. Additionally, the mission is condensed into critical waypoints, which fully define the decision process and allow the heuristic layer to be designed with understandable parameters and crisp logic. Risk assessment is performed by applying Bayesian regression to vehicle behavior metrics, which provides computationally efficient equations and expressions to the MPC.

\subsection{Related work}

Several papers have considered risk measures in planning and handling uncertainties in an MPC framework, as summarized in \cite{mayne,mesbah}, and shown in \cite{cautiousmpc,pavone,riskbook,riskmeasures,park}. In \cite{cautiousmpc}, the authors study uncertainty propagation to ensure chance constraints on a race car; results show that the algorithm can learn uncertainty in the dynamics, associate risk to the unknown dynamics, and plan so that the trajectories are safe. The algorithm is efficient, but the prediction horizon is relatively short, and some approximation of the dynamics was necessary for real-time performance. In \cite{pavone}, the authors provide stability proofs for a linear MPC controller, which minimizes time-consistent risk metrics in a convex optimization form. These papers focus on operating in a constrained environment or under controlled assumptions to provide uniform guarantees. Our work's key difference is that in this paper we relinquish online risk constraint satisfaction to external heuristics, widening the solver's capabilities and flexibility at the cost of a more conservative solution. Apart from fundamental results in this field, such as \cite{whittle}, modern developments  in \cite{Sopasakis} show that the increased computational capacity is enabling risk-minimization to be executed in real-time for a variety of systems. In \cite{Sopasakis}, the authors provide a synthesis methodology for risk-averse MPC controllers for constrained nonlinear Markovian switching systems.

Few papers have been published concerning highly stochastic rendezvous problems. Most notably, in \cite{rucco} the authors compute optimal trajectories in refueling missions, but in their work most of the uncertainty is environmental and local, whereas we consider epistemic and large-scale uncertainties. Additionally, because of the large-scale, we focus efforts on efficiency and scalability instead of the accuracy of dynamic and disturbance models. Risk minimization in optimal control has been studied in several papers \cite{cautiousmpc,pavone,whittle,Sopasakis,fleming}. Minimizing risk in optimal control is traditionally intractable \cite{cautiousmpc,pavone} for real-time implementation, a capability which the proposed method is designed to have.

\subsection{Statement of contributions}

We present a hybrid algorithmic MPC framework to solve the running rendezvous problem in real-time under large uncertainties. We aim to condense a large-scale optimization problem into a few critical variables that minimize total time and energy consumption under the non-negligible probability of mission failure, which is handled by a robust and fast heuristic layer.

Structurally, a Bayesian learning component approximates driver behavior while enabling risk bounds to be efficiently computable. Parameterization of the path and velocities allows the data-driven Bayesian learner to remain fast, which, coupled with the MPC controller's low dimensionality, is shown to run in real-time even under highly nonlinear constraints. No approximation is made in the OCP itself; the solution is locally near-optimal up to the learned model quality.

We show that our method is flexible and robust, where a considerable portion of the computational complexity can be executed apriori. We provide a scenario that illustrates the advantage of this approach, showing both successful and unsuccessful outcomes. 

The rest of this paper is structured as follows: in Section \ref{sec:problem}, we introduce and define the problem in algorithmic format. In Section \ref{sec:meth}, we present the two main components of this approach: the model learning and risk estimator, and the optimization problem statement. In section \ref{sec:res}, we demonstrate two example scenarios showing the decision making aspect of the algorithm. Finally, in Section~\ref{sec:conc}, we provide concluding remarks and discuss the shortfalls of this approach and future directions to address them, respectively.

\section{Problem Formulation}
\label{sec:problem}

We begin by defining the notion of persistent safety.
\newtheorem{definition}{Definition}
\begin{definition}[Persistent Safety]
\label{def:persafe}Let $x_{k+1}=f(x_k,u_k)$ be a system with state vector $x\in\R^n$ and control vector $u\in\R^m$. A safety set $S_k\subset(X,U)$ is a set, in which all states and inputs are considered safe by some measure $\rho(x):\R^n\rightarrow\R$ at step $k$. We define a planning algorithm as persistently safe, if $S_k = \{x_k\in X,\;u_k \in U : f(x_k,u_k) \in S_{k+1}\}$ exists for all $k$ for a set of admissible states $X$ and control inputs $U$.
\end{definition}

The goal is to compute a persistently safe trajectory (sequence of states $x$ and inputs $u$) as defined in Definition \ref{def:persafe} that satisfies a rendezvous condition. This is achieved by postponing a decision between aborting or continuing the mission for as long as possible. The additional time afforded by postponing this decision is used to improve uncertainty prediction and, consequently, reducing the risk of running out of battery or fuel. 

\begin{figure}
    \centering
    \includegraphics[width=0.40\textwidth,trim=180mm 100mm 180mm 75mm, clip]{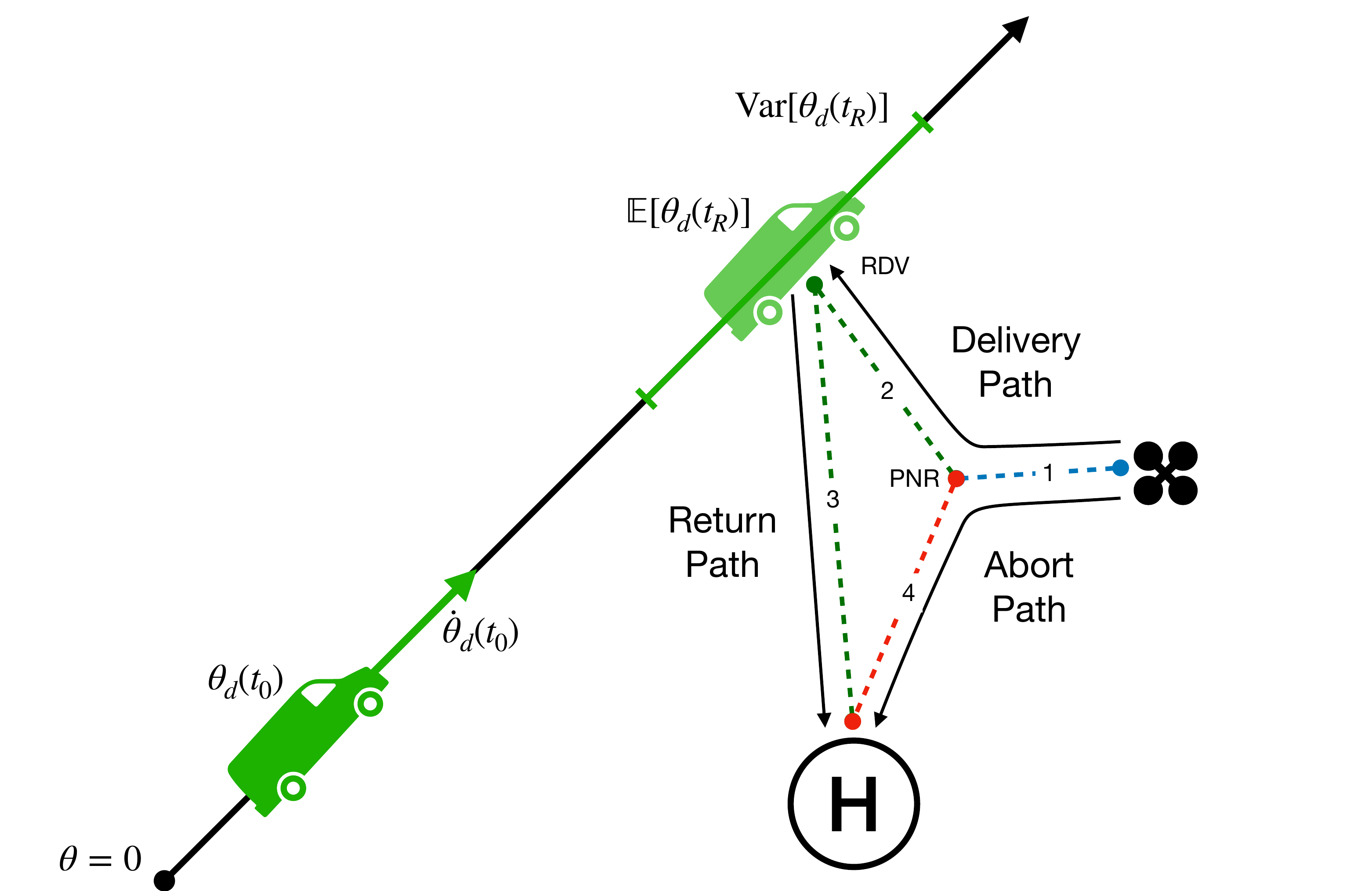}
    \caption{Overview of the problem setup at  time instance $t_0$. Two separate paths are computed in parallel; one to rendezvous and with the ground vehicle and return, and another to abort. $\E[ \theta_d(t_R) ]$ and $\Var[ \theta_d(t_R) ]$ indicate the randomness associated with $\theta_d(t_R)$; uncertainty in driver behavior and inclusion of sensor noise imply that this prediction is a random variable.}
    \label{fig:overview}
\end{figure}

For this problem, a parameterized path $p(\theta),\;p:\R^+ \rightarrow \mathbb{R}^2$, $\theta\in\R^+$, and historical velocity data along the path $\dot{\theta}_h(t)$, $\td_h:\R^+\rightarrow\R$ obtained from the traffic data are provided apriori. A stream of noisy position $\theta_d(t)$, $\theta_d:\R^+\rightarrow\R^+$, and velocity $\td_d(t)$ measurements from a driver moving along the path are obtained in real-time via on-board sensors. We wish to find a rendezvous point $\theta_d(t_R)$ that brings both vehicles together at a rendezvous time $t_R\in\R^+$. Due to sensor noise and uncertain driver behavior, we aim to estimate the distribution of $\theta_d(t_R)$ and plan on it. Along with the rendezvous point, we also determine a Point-of-No-Return (PNR) between the UAS and $\theta_d(t_R)$, from which a separate path navigates the UAS  to a safe landing location in case the risk of rendezvous failure $\rho(\theta_d(t_R))$ that maps the distribution of $\theta_d(t_R)$  to $\R^+$ \cite{riskmeasures} is too great. We model the UAS with single integrator dynamics in this context. We define safety (and, thus, its associated risks) as a function of the probability of running out of remaining battery or fuel $E_r$. Figure \ref{fig:overview} illustrates the setup.

Algorithmically, we maximize the amount of data gathered before the PNR is reached and check if that is sufficient to guarantee a low-risk rendezvous. The OCP constraints ensure that the decision time will tend to zero as $\rho$ increases, and time progresses. When this occurs, the heuristics evaluate a risk measure and commit to a route via predetermined thresholds. With this structure, the MPC is agnostic to the risk measure.

The MPC minimizes the risk associated with the rendezvous location itself ($\rho_R$), as well as the rendezvous time $t_r$. As the process is initiated, the lack of data will lead to excessively large risks; thus, it is beneficial to prolong the mission and maximize data gathering via a decision time $t_1$ between the UAS and the PNR. The stop condition is $t_1\leq\epsilon$. This condition indicates the moment where the decision time is at its allowable minimum, which eventually occurs given the problem's spatial constraints. Once this threshold is met, we check $\rho\leq E_\mathrm{risk}^\mathrm{max}$, where $E_\mathrm{risk}^\mathrm{max}$ is defined as the maximum energy needed to complete the rendezvous trajectory inside the confidence bound $\gamma_\mathrm{max}$ of $\theta_d(t_r)$.

As the ground vehicle travels along $p(\theta)$, we collect data and append it to a dataset $\mathcal{D}$ containing the driver's velocity $\td_d$ and expected velocity $\td_h$. This dataset is then used to produce mean $\mu_w$ and variance $\Sigma_w$ functions of the driver's position in the future, a process described in Section \ref{sec:bayes}. The mission algorithm does this process iteratively, sampling new data to improve the driver model and at the same time running an MPC loop to plan the routes $(\bv,\bx,\bt)$, defined in Section \ref{sec:mpc}. If either $\rho_A$ grows above a threshold $\gamma_A$ or the decision time $t_1$ falls to $\epsilon$, the risk measures are evaluated and a commitment is made. This process is described in Algorithm \ref{al:main}. Notice that traditional MPC approaches involve a moving finite time horizon, which is updated at every time step. Here the time horizon is fixed, but we update how that time is divided at each time step. Because the mission is re-planned the same way, we classify this controller as an MPC-type controller.
\begin{algorithm}[h]
\SetAlgoLined
$\mathcal{D}\leftarrow$ Initial Data\\
 \While{$t_1>\epsilon$}{
    $\mu_w,\;\Sigma_w \leftarrow \text{Regress}(\mathcal{D})$\\
 	$\bv,\bx,\bt\leftarrow \text{MPC}(\mu_w,\Sigma_w,\bx)$\\
 	Send Control Input $\bv$ to UAS\\
 	$\mathcal{D}\leftarrow \text{Append(New Data},\mathcal{D})$}
  \uIf{$\rho(\mu_w,\Sigma_w,\bx)\leq E_\mathrm{risk}^\mathrm{max}$}{
    Proceed with rendezvous
  }
  \Else{
    Abort and return
  }
 \caption{Mission Algorithm}
 \label{al:main}
\end{algorithm}
\subsection{Problem Novelty}

The problem of performing a rendezvous with (or intercepting) a moving target is not new. However, these problems fall into two distinct categories: interception of a target on a known path or interception of a target with an unknown trajectory \cite{pursuitevasion}. The interception of a target following a known path is seen as trivial. With full knowledge of the target behavior, infinitely many trajectories will intercept the target at a chosen time \cite{ethtraj}. In these scenarios, the goal is to find the \textit{optimal} trajectory for the interception. Added uncertainties such as model discrepancies and environmental disturbances are often included, but do not change the architecture.

The opposite problem is the one that intercepts a target following an unknown path. Depending on the objective and constraints, this scenario is significantly more challenging. Robust solutions such as constant line-of-sight angle laws \cite{robotlos,missle} can guarantee interception at some point in time, but do not satisfy optimality. Modern solutions such as trajectory prediction schemes \cite{ethtraj} aim to predict the target trajectory to then plan an intercept course.

In this paper, we argue that the problem depicted in Figure \ref{fig:overview} does not fall in either category, thus requiring a custom solution. Although we know that the target's spatial constraints (ground vehicle) are determined by the road it is traveling on, we do not know its temporal trajectory. This, combined with the large scale of the problem and its risk constraints, makes the available tools unsuitable. Feedback control laws cannot account for these same constraints, and, at this scale, for this application, spatial trajectory prediction is computationally hard to execute and unnecessary given our geometric knowledge.

\section{Methods}
\label{sec:meth}

In this section we discuss the components of Algorithm~\ref{al:main}. First, we define the learning component based on Bayesian linear regression. This regression approximates driver behavior and provides mean and variance functions for the vehicle's future location. Next, we outline the solver layer in the form of an MPC controller, which plans a rendezvous location given the driver behavior estimate. Finally, we briefly discuss risk measures and dangers when implementing them in the proposed framework.

\subsection{Bayesian Linear Regression and Risk Assessment}
\label{sec:bayes}

In this section we discuss the Bayesian learning component introduced in Section \ref{sec:problem} and represented in Algorithm \ref{al:main} as the regression function. One of the major challenges for the proposed problem is that each driver behaves differently. While one driver may drive at a conservative speed limit, another might drive relatively faster. Therefore, learning a driver's `behavior' will be beneficial to the rendezvous problem. We now set up this learning problem. Consider the parameterized path $p(\theta)$, $\theta \in \R^+$. We assume that we have access to the driver's position $\theta_{d,i} = \theta(t_i)$, where $t_i$ is the time instance, at which the measurement is obtained. Furthermore, we have measurements of the driver's velocity denoted by $\dot{\theta}_{d,i} = \dot{\theta}_d(t_i)$. All measurements are considered to have additive noise. We also assume that we have access to historical velocity profile given by $\dot{\theta}_{h,i} = \dot{\theta}_h(t_i)$. Such a historical velocity profile can be generated by collecting measurements of vehicles traversing the path $p(\theta)$ and fitting a distribution over it using methods similar to those in \cite{traffic1,traffic2}. In our case, we assume the historical velocity profiles are in the form of a Gaussian distribution (explained later). To summarize, given the driver's position $\theta_i$, we have access to a measurement of the driver's velocity $\dot{\theta}_{d,i}$ and the corresponding probabilistic historical velocity $\dot{\theta}_{h,i}$. A comparison of $\dot{\theta}_{d,i}$ and $\dot{\theta}_{h,i}$ thus represents a measure of the driver's behavior. In particular, we wish to learn $\dot{\theta}_d(\dot{\theta}_h): \mathbb{R} \rightarrow \mathbb{R}$.

The traditional approach would be to directly learn the vehicle's position function $\theta_d(t)$; however, this would cause the uncertainty propagation to expand too quickly and force an abort decision too often \cite{cautiousmpc}. Instead, we explore both the fact that the vehicle is constrained to a known path and that the velocity along the path has a strong prior (the historical velocity $\dot{\theta}_h(\cdot)$). A disadvantage of this approach is that an integration procedure must be carried out to estimate $\theta_d(t)$. In a regular OCP formulation, this function would be forward-Euler integrated inside the solver in the form of dynamic model constraints \cite{borrelli}. However, due to the coarse discretization considered in this paper, such implementation would not be feasible. Instead, we use a finite basis model that enables analytical integration to be performed offline.

   \begin{figure}
      \centering
      \includegraphics[trim=0mm 58mm 0mm 15mm, clip, width=0.38\textwidth]{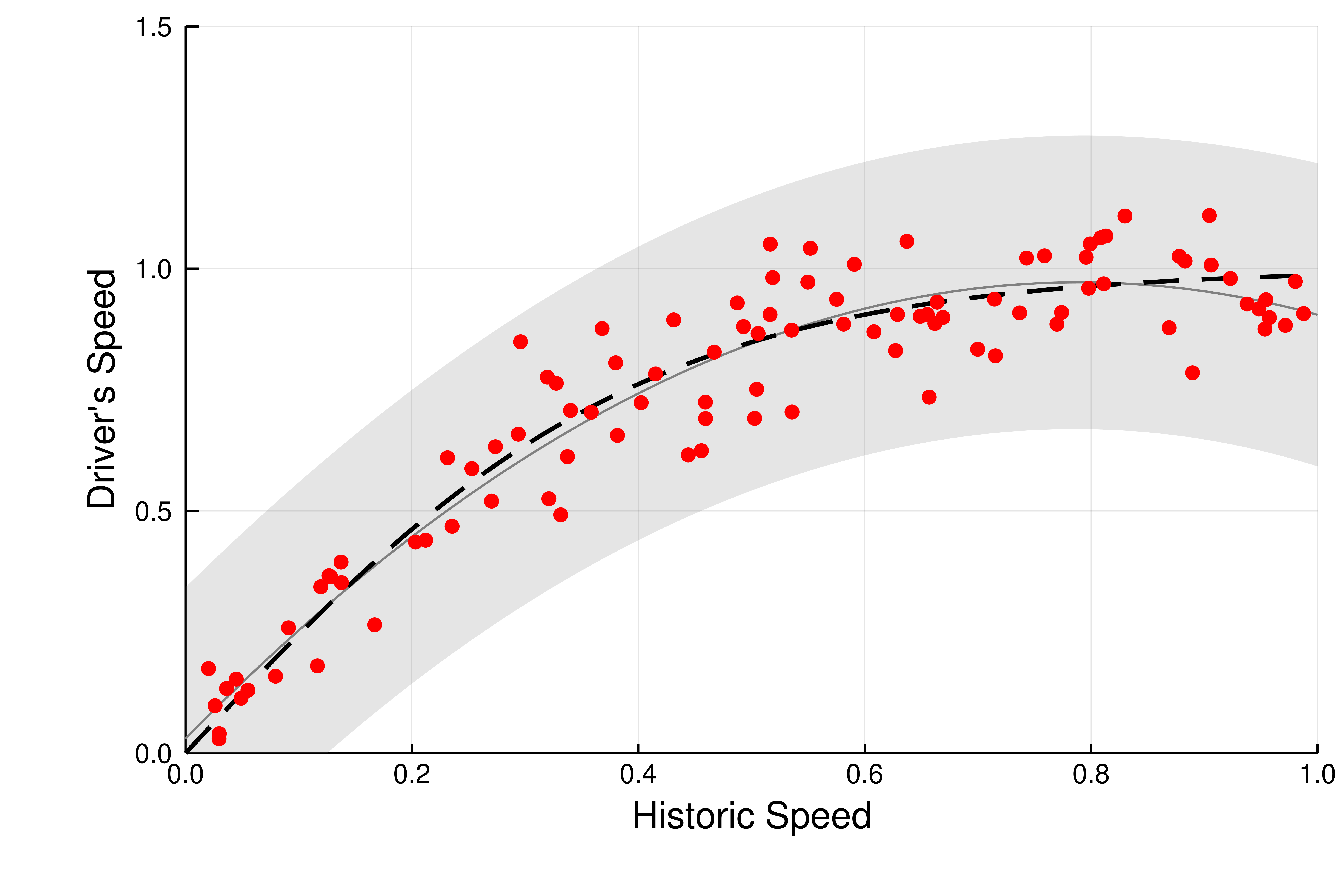}
      \caption{Example regression: a nonlinear curve we wish to learn (dashed) maps historical data to driver velocities, by regressing on the measurements we get: mean (gray line) and variance functions (grey shaded area).}
      \label{fig:fitting}
   \end{figure}
   
We assume that the mapping $\dot{\theta}_d(\dot{\theta}_h)$ admits a linear model with a finite number of basis functions as $\td_d(\td_h) = w^\top \phi(\td_h),\; w \in \mathbb{R}^m$, 
where $\phi:\mathbb{R} \rightarrow \mathbb{R}^m$ is the vector of known basis functions and $w \in \mathbb{R}^m$ are the weights to be learned. We place the following prior on the weight vector $w$ as 
\begin{equation}\label{eqn:prior}
w \sim \mathcal{N}(0_m,\Sigma_m), 
\end{equation}
where $\Sigma_m \in \mathbb{S}^m$ is the covariance obtained using historical data. Note that we assume a zero prior mean without  loss of generality. As we will see, since we consider noisy measurements corrupted by zero-mean Gaussian noise, non-zero prior mean can be easily incorporated~\cite[Sec.~2.7]{williams2006gaussian}. 
We assume a noisy data stream of the form
\begin{equation}\label{eqn:measurements}
    y_i = \td_d(\td_{h,i}) + \zeta, \quad \zeta \sim \mathcal{N}(0,\sigma^2), \quad i \in \{1,\dots,N\}.
\end{equation} \textcolor{black}{ Considering the fact that by construction we have a finite-basis linear model and Gaussian measurements, we can construct the posterior distribution by conditioning the prior in~\eqref{eqn:prior} on the measurements in~\eqref{eqn:measurements}. This is known as Bayesian Linear Regression~\cite[Sec.~2.1]{williams2006gaussian}~\cite[Sec.~3.3]{bishop2006pattern}. For the case of a linear model with finite-number of basis functions, Bayesian Linear Regression (BLR) is equivalent to Gaussian Process Regression (GPR) with the kernel function induced by the basis functions $\phi(\cdot)$~\cite[Sec.~2.1]{williams2006gaussian}.
Then the natural question arises regarding the use of BLR and not GPR to accomplish the desired goals since the apriori choice of a finite number of basis limits the models' expressive flexibility. A straightforward argument is that BLR is computationally cheap compared to GPR, especially as a function of available data. This is crucial considering the online nature of the proposed method. Furthermore, one can always approximate the well-known stationary kernels like the Squared-Exponential (SE) or Mat\'{e}rn kernels using a finite number of random Fourier features~\cite{rahimi2008random}. Finally, we would like to highlight the fact that any estimation method which provides a notion of uncertainty can be used since the algorithm is agnostic to the risk estimation layer as it will be shown in subsection \ref{sec:mpc}.}

At any given time, the algorithm can sample a position $\theta_d$ and velocity $\td_d$ of the vehicle. Then using the prototypical velocity profile $\dot{\theta}_h$, we can generate the data in $\mathcal{D}$. We write the data in compact form as $\mathcal{D} = \{D,H\}$, where $D,~H \in \mathbb{R}^N$ are defined as
\[
D = \begin{bmatrix} \td_{d,1} & \cdots & \td_{d,N} \end{bmatrix}^\top, \quad 
H = \begin{bmatrix} \td_{h,1} & \cdots & \td_{h,N} \end{bmatrix}^\top.
\]

Given the measurements~\eqref{eqn:measurements} and the prior~\eqref{eqn:prior}, we obtain the posterior distribution of the parameter vector $w$ as 
\begin{align}
\label{eqn:posterior} &w \in \mathcal{N}(\mu_w,\Sigma_w),
\end{align}
where $\quad\mu_w = \frac{1}{\sigma^2}A^{-1} \Phi(H) D,\; \Sigma_w = A^{-1}$, and  $A = \sigma^{-2}\Phi(H)\Phi(H)^\top + \Sigma_m^{-1}$.

With the mean and variance functions fitted, the next necessary step is to forward propagate these functions with respect to the historical data. The challenge is that the model represents a mapping between velocities, with no spatial information otherwise. Because we used parameterized velocities instead of the Euclidean representation,  it is possible to integrate along the path using the known velocity profile from historic data and the path information itself. Suppose that at some time instance $t_0$ the rendezvous vehicle is at $\theta_{d,0} = \theta_d(t_0)$, and we wish to estimate the vehicle's position at some instant $t_f > t_0$. Given the integrable temporal prototypical velocity profile $\td_h(t)$, the predictive distribution of $\theta_d(t_f)$ can be computed via
\begin{align}
    \theta_d(t_f) = \theta_{d,0} + \int_{t_0}^{t_f}(\td_d(\td_h(\tau))\diff{\tau} \label{eqn:posterior_position} = \theta_{d,0} + w^\top \psi(t_f),
\end{align} where $\psi(t_f) = \int_{t_0}^{t_f}\phi(\td_h(\tau))\diff{\tau}$. This integral can be computed, as a function of $t_f$, apriori. For example, for polynomial kernels or the random Fourier feature approximation of SE or Matern kernels, using the posterior distribution of $w$ in \eqref{eqn:posterior}, and the fact that $\theta_d(t_f)$ is a linear transformation of the random variable $w \in \mathbb{R}^m$, we get the following posterior distribution
\begin{align}
    \mathbb{E}[\theta_d(t_f)] = & \theta_{d,0} + \mu_w^\top \psi(t_f),\\
    \text{Var}[\theta_d(t_f)] = & \psi(t_f)^\top A^{-1}\psi(t_f). 
\end{align} As the order of the basis increases, and depending on the structure of the historical velocity profile, the explicit form can become cumbersome. However, it is all done apriori and automated, and because the range of velocities we expect to encounter is small, we generally do not need a large number of basis.

\subsection{MPC formulation}\label{sec:mpc}

In this section we discuss the structure and particulars of the MPC component introduced in Algorithm~\ref{al:main}. A primary challenge of the rendezvous problem is presented by the trajectories under strict and numerous constraints, of which many are non-convex. By exploring two special features of the problem formulation we reduce dimensionality and attain tractability. We now outline the Optimal Control Problem (OCP) associated with the rendezvous problem. As mentioned previously in Sec.~\ref{sec:problem}, the solver is tasked with finding two critical points: (Point-of-No-Return) PNR and the rendezvous location $p(\theta_d(T_R))$. To fully define the problem and gain temporal constraint management we expand the control from velocities to also include a time ``input''. The nature of this problem requires the UAS to coincide with the vehicle both in space and time. By introducing time as a manipulated variable in the OCP, we allow for the solver to directly decide on the optimal time and place for the rendezvous maneuver to occur. This time input works by assuming a piece-wise constant control law along each of the four segments (PNR, rendezvous, landing location, and abort location), which is  possible due to our assumption on the UAS integrator dynamics described below:
\begin{align}
    x_k &= x_{k-1} + v_kT_s\\
    E_{r,k} &= E_{r,k-1} - \left(\dfrac{mv^2}{2} + \alpha m\right) T_s,
\end{align}
where $T_s$ is the sampling time, $m$ the scalar vehicle mass, and $\alpha$ the scalar hovering energy consumption constant. We represent each of the segments using the state vector $(\bx,\bv,\bt)\equiv(x_i,v_i,t_i)$, $i\in\{1,\dots,4\}$. Here, $t_i$ represents the time to be spent at a constant velocity $v_i$ to reach one waypoint from another. Furthermore, $x_i\in\R^2$ represents each of the defined physical waypoints in Euclidean coordinates and $v_i\in\R^2$ represents velocity inputs in Euclidean coordinates. The waypoints are, in this order, the Point-of-no-Return (PNR), the rendezvous location (RDV), the landing location, and the abort location, as shown in Figure \ref{fig:overview}. The designed Optimal Control Problem (OCP) is given by:
\begin{subequations}\label{eq:ocp}
\begin{alignat}{3}
    \min_{U,T_R}\quad   & \rho_R(T_R,\Sigma,\bx) + \left(\sum_{i=2}^3t_i - t_1\right)                            && \\
    \text{s.t.}\quad    & x_i = x_{i-1} + v_it_i,~x_4 = x_1 + v_4t_4,   \label{eq:dyn}&& \\
                        & |v_i| \leq v_\mathrm{max},                      \label{eq:vel}&& \\
                        & x_3 = \E[\theta_d(T_R)],~ x_4 = S_L,~x_5 = S_A,\label{eq:abp}&& \\
                        & \sum_{i=1}^3 t_i \leq t_\mathrm{max},~t_1 + t_4 \leq t_\text{max},~t_c \leq t_i                                        \label{eq:tim}&& \\
                        & E_1 + E_2 + E_3 \leq E_r,~E_1 + E_4 \leq E_r, \label{eq:er1}
\end{alignat}
\end{subequations}
where $T_R \equiv t_1+t_2$, $S_L$ and $S_A$ are the landing and abort destinations, $E_r$ the remaining energy $E_i$, $t_c$ a dwell time for the low level controller to switch tracked segments, $\Sigma$ is a variance function associated with the mission state given by \eqref{eqn:posterior}, and $\rho_R$ any risk measure we wish to minimize. The dwell time is necessary to stop the solver from placing waypoints arbitrarily close to each other and creating undesirable sharp turns, which are problematic for our single integrator dynamics assumption. Moreover, $U\equiv\{t_i,v_i\},\;i\in(1,...,4)$. Formulating this problem with risk constraints instead of cost would cause the solver to potentially deny dangerous solutions instead of postponing a decision and waiting for new data that can eventually yield a feasible solution. Since risk constraints still need to exist, their satisfaction is relegated to the heuristics discussed in Algorithm \ref{al:main}.

All of these constraints are natural because every single one is predetermined at the design stage. For example, constraint \eqref{eq:er1} is directly produced from the battery used in the UAS, and constraint \eqref{eq:abp} is given from the map where the mission takes place. The only task left for the designer is to choose the risk measures. Fortunately, because the proposed method is agnostic to risk measures, the designer can choose with no concern over tractability or internal conflicts in the solver.

\begin{figure}
    \centering
    \includegraphics[width=0.40\textwidth,trim=180mm 100mm 180mm 75mm, clip]{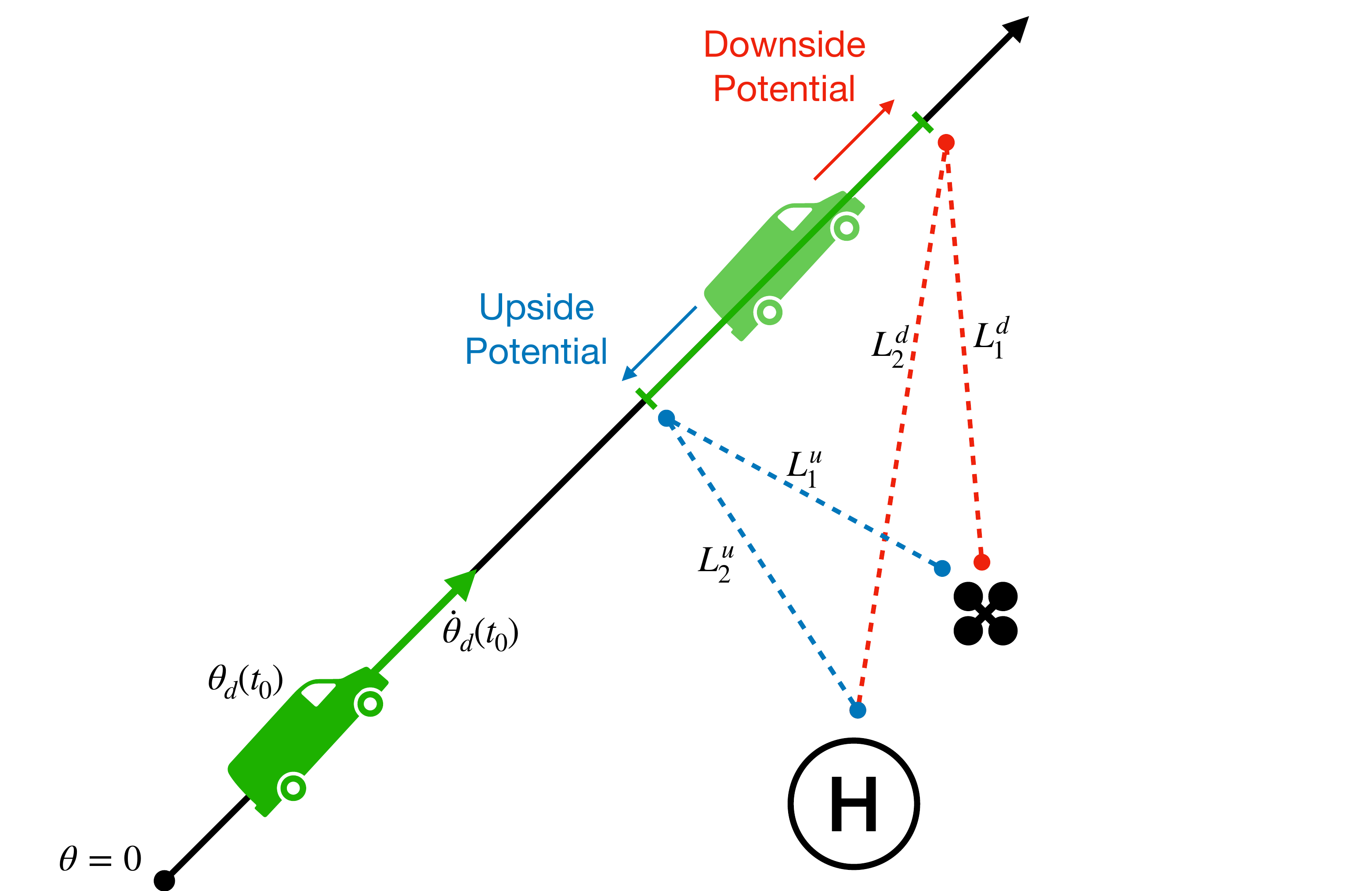}
    \caption{Downside Risk measure: the two outcomes within some confidence bound at time $t_R$. The downside potential consumes more power because $L^d_1+L^d_2>L^u_1+L^u_2$. The Downside Potential Energy is the extra energy necessary should the red outcome occur. In this example, $\rho=E_1+E_2+E_3-(E_{L^d_1}+E_{L^d_2})$.}
    \label{fig:dsrisk}
\end{figure}

Quantifying risk is the effort of determining a measure $\rho$ that maps a set of random variables to a real number \cite{riskmeasures,riskbook}. With this definition, the random variables are the states of the UAS (due to process and measurement noises) and, more importantly, the position of the ground vehicle due to the driver's uncertain behavior. It is crucial to choose measures that reflect meaningful quantities in the problem formulation. In this framework, the risk is directly related to the uncertainty regarding the vehicle's location in the future and the limitations that the path imposes on planning. If the driver is erratic, or the path only allows the rendezvous to happen in unfavorable locations, we consider that the mission has elevated risk. Several risk measures are popular; some examples are Expectation-Variance \cite{whittle}, (Conditional, Tail) Value-at-Risk  ~\cite[Sec.~3.3]{riskbook}, and Downside Variance ~\cite[Sec.~3.2.7]{riskbook}. These measures can introduce nonlinearity and preclude gradient information, endangering tractability. A popular approach uses gradient-free methods, which sample these measures and choose inputs corresponding to minimum risk \cite{samprisk}. In this paper, we tackle tractability by exploring the problem geometry to its fullest and reducing the number of variables to an absolute minimum, as shown above, and in Section \ref{sec:bayes}. In this paper, we consider two risk measures. Measure $\rho$ is the Downside Potential \cite{riskbook} of energy expenditure. It computes the worst outcome energy-wise as the maximum energy necessary to complete the mission within the driver position confidence bounds. Furthermore, $\rho_R$ is defined as the variance of driver position at time $t_R$. These measures provide guarantees that the UAS will return and favor locations on the road where uncertainty is minimal, respectively. Figure \ref{fig:dsrisk} depicts this setup, where one tail of the distribution is a downside tail, and the other an upside tail.

\section{Results}
\label{sec:res}
\urlstyle{tt}

This section presents two scenarios with distinct outcomes. In the first scenario, the mission is successful, since when the decision time $t_1$ reaches $\epsilon=5$ seconds, the Downside Energy Potential $\rho$ is below its threshold of $200\si{\joule}$. In the second scenario, we increase noise and make the driver erratic, which increases uncertainty makes the UAS abort a risky rendezvous in favor of returning home. To simulate the dynamics, we use an Euler integration scheme with a discretization step of one second. The OCP solver reaches a solution in a median time of 75.604ms on a 2012 3.4 GHz Quad-Core Intel Core i7 implemented in Julia with the Ipopt solver. At this rate, the algorithm runs faster than the discretization time. The source code can be found at \url{https://github.com/gbarsih/Safe-Optimal-Rendezvous}. The mission takes place on a 1\si{\kilo\metre^2} area, with a stretch of road going from (0,0)\si{m} to (1000,1000)\si{m}, as shown in Figure \ref{fig:overview}. We set the abort, landing, and take-off locations ($S_A$, $S_L$, and $x_0$, respectively) to the same position at $(500,0)\si{m}$, the path to $p(\theta)=(\theta,\theta)\si{m}$, and driver behavior functions to $\td_d(\td_h)=a\td_h(t)$, so that $a>1$ makes a driver proportionally faster than the average driver. On that same path we set $\td_h(t)=10(1- t/200)$ and $\rho_R(\theta_d)=\mathrm{Var}[\theta_d(t_R)]$.

\subsection{Maximizing Decision Time}

\begin{figure}
    \centering
    \includegraphics[width=0.42\textwidth]{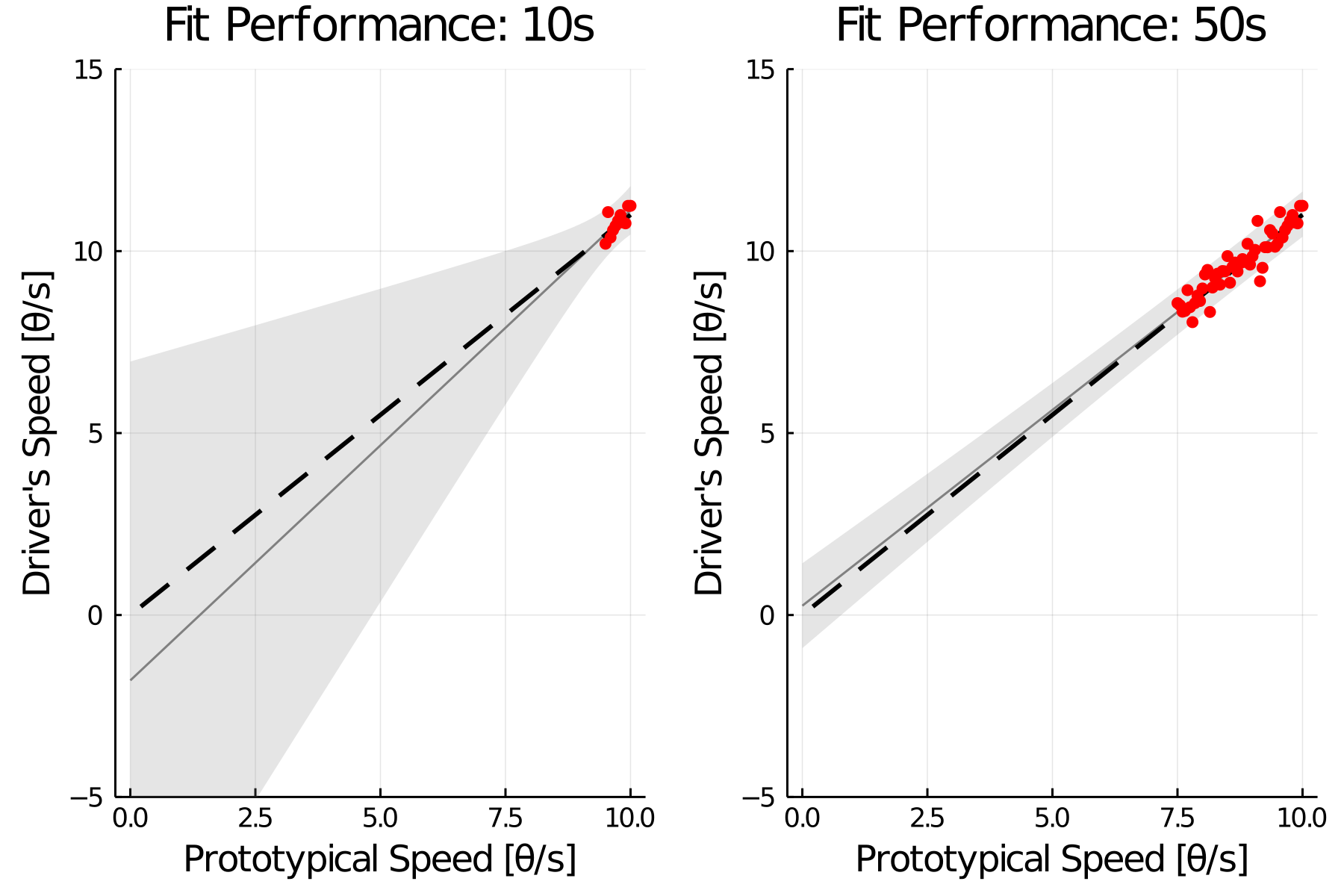}
    \caption{Bayesian fit performance at two points in time: more data completes the driver's behavior profile. Waiting for more data means noise is averaged, and we explore more speeds.}
    \label{fig:fitperf}
\end{figure}

One of the algorithm's core functionalities relies on postponing a decision by maximizing $t_1$ to gather more data and attempt to find a safe trajectory. This maximization goes against the mission objective of minimizing time but is necessary to guarantee safety and feasibility. Figure \ref{fig:fitperf} shows exactly that, for $a=1.1$, where more data at 50 seconds reduces uncertainty in unknown speeds.
\subsection{ Low-Risk Mission}

\begin{figure}
    \centering
    \includegraphics[width=0.42\textwidth]{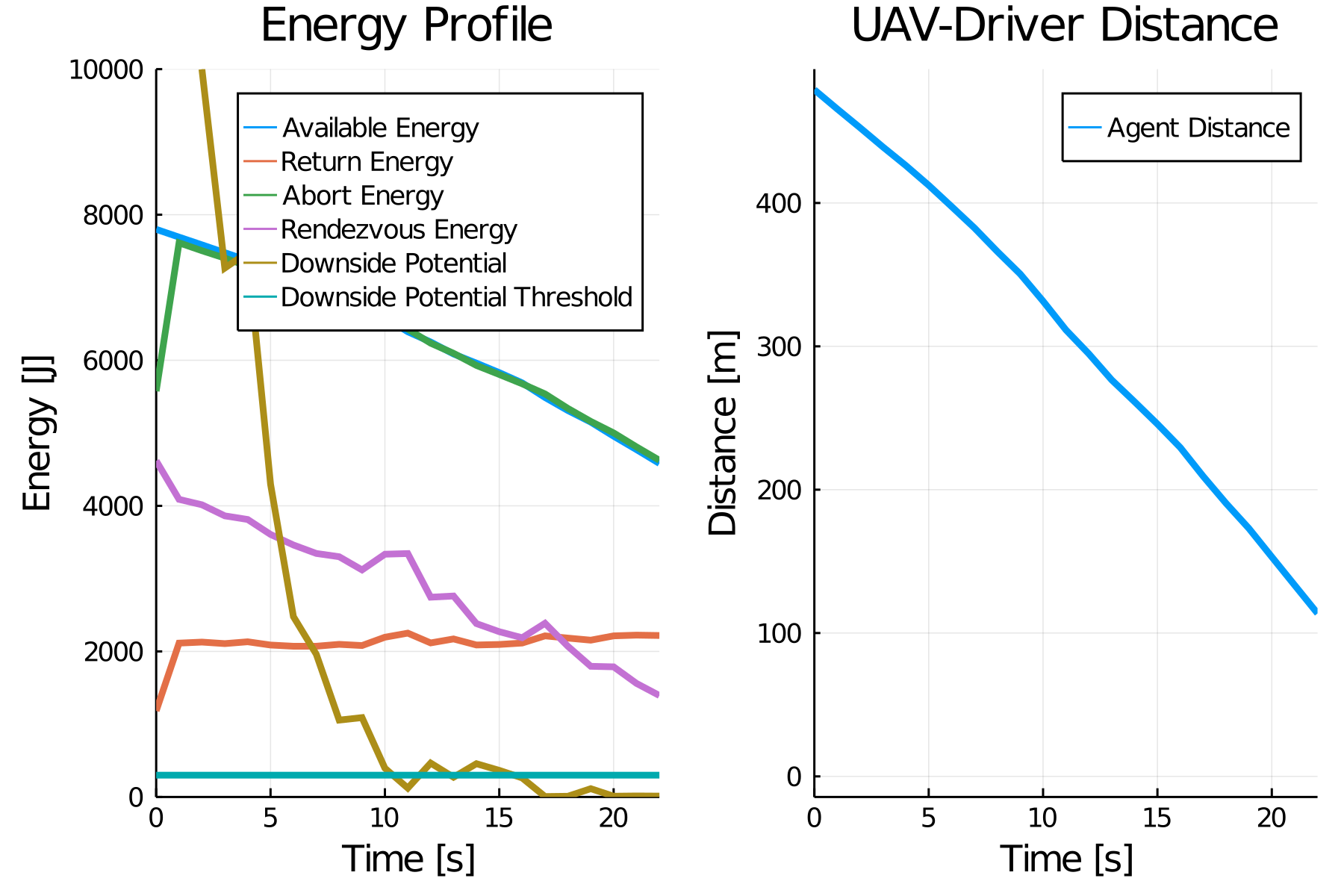}
    \caption{Outcome of a low-risk decision: in 20 seconds available to make a decision, enough data is collected to assert high confidence of success.}
    \label{fig:good}
\end{figure}

In this example, we set $\sigma=3$ and $a=1.1$. In 20 seconds, the algorithm terminates and is forced to make a decision. We notice that the abort energy is always maximized up to the available energy and that the downside potential falls below the threshold despite being prohibitively high in the early stages. Figure \ref{fig:good} shows the energy profile of each trajectory segment, available energy, and downside potential energy, as well as the distance between the car and the UAS for the low-risk scenario.
\subsection{High-Risk Mission}

\begin{figure}
    \centering
    \includegraphics[width=0.42\textwidth]{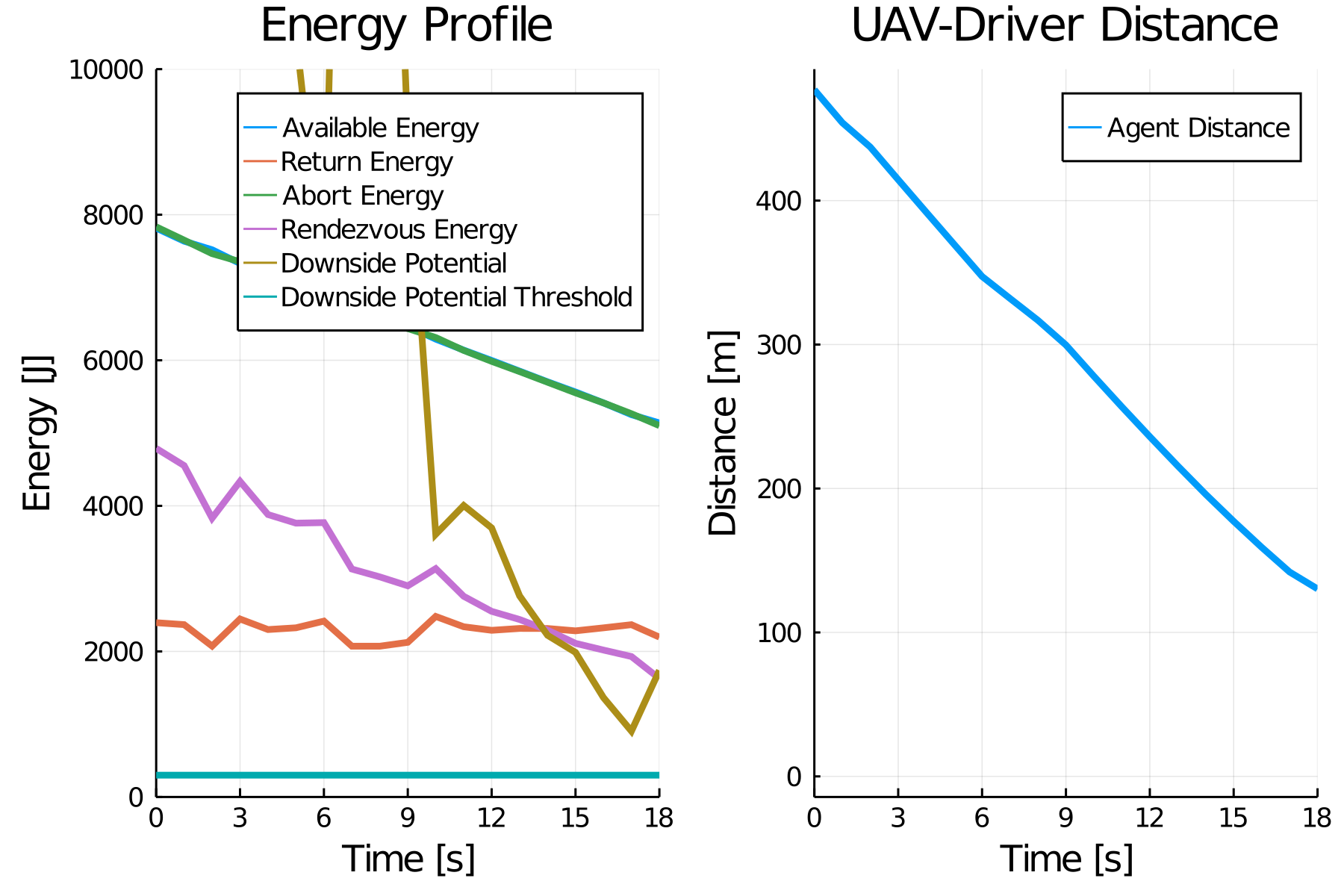}
    \caption{Outcome of a high-risk decision: in 18 seconds, not enough data was collected, or it was, and risk is too high. At decision time, the downside potential is above the threshold, and a decision is made to abort the mission.}
    \label{fig:bad}
\end{figure}

In this example, we set $\sigma = 6$ and $a=1.3$, meaning that the sensor has more noise and the driver is more erratic. In 18 seconds, the algorithm terminates and is forced to make a decision. Because of the heightened uncertainty, the downside potential is high, and the UAS has a high probability of running out of battery. In this case, the algorithm will trigger the abort decision, which is feasible because that path was planned. Figure \ref{fig:bad} shows the energy profile of each trajectory segment, available energy, and downside potential energy, as well as the distance between the car and the UAS for the high-risk scenario.
\section{CONCLUSIONS}
\label{sec:conc}

We presented an algorithm capable of planning a long-distance rendezvous between an autonomous aerial vehicle and a ground vehicle traversing a path. The large uncertainties associated with driver behavior combined with finite energy to complete the mission require risk management, a learning component, and an MPC-like controller to work in unison and guarantee safety at all times, even if safety translates to aborting the mission. The algorithm is shown to be persistently safe due to the continuous planning of an abort path at all times. For future work, we intend to expand this work in two ways. First, a modification of this algorithm to accomodate pruing paths. Second, a parcel delivery framework like the one described here will have multiple vehicles in the logistic matrix. A third decision path can be added by leveraging this, which waits for a new ground vehicle to enter the matrix and then deliver to that vehicle.




\bibliographystyle{IEEEtran}
\bibliography{bibliography.bib}

\end{document}